\documentclass[aps,prl,amsmath,twocolumn,groupedaddress]{revtex4-1}
\usepackage[english]{babel}
\usepackage{CJKutf8}
\usepackage{graphicx}
\usepackage{subfigure}
\usepackage{dcolumn}
\usepackage{bm}
\usepackage{units}
\usepackage{siunitx}
\usepackage{color}
\usepackage{longtable}
\usepackage{amsmath}
\usepackage{amssymb}
\usepackage{array}
\usepackage{multirow}
\usepackage{tablefootnote}
\usepackage[letterpaper,total={7in,9.5in},top=0.75in,left=0.75in]{geometry}
\usepackage[colorlinks=true,linkcolor=blue,citecolor=blue,urlcolor=blue,filecolor=blue]{hyperref}

\newcolumntype{L}[1]{>{\raggedright\let\newline\\\arraybackslash\hspace{0pt}}m{#1}}
\newcolumntype{C}[1]{>{\centering\let\newline\\\arraybackslash\hspace{0pt}}m{#1}}
\newcolumntype{R}[1]{>{\raggedleft\let\newline\\\arraybackslash\hspace{0pt}}m{#1}}

\begin{document}

\title{Sisyphus Optical Lattice Decelerator}

\author{Chun-Chia Chen (陳俊嘉), Shayne Bennetts, Rodrigo Gonz\'alez Escudero, Florian Schreck, Benjamin Pasquiou}
\email[]{SOLD@strontiumBEC.com}
\affiliation{Van der Waals-Zeeman Institute, Institute of Physics, University of Amsterdam, Science Park 904,
1098XH Amsterdam, The Netherlands}

\date{\today}

\begin{abstract}

We experimentally demonstrate a variation on a Sisyphus cooling technique that was proposed for cooling antihydrogen. In our implementation, atoms are selectively excited to an electronic state whose energy is spatially modulated by an optical lattice, and the ensuing spontaneous decay completes one Sisyphus cooling cycle. We characterize the cooling efficiency of this technique on a continuous beam of Sr, and compare it with radiation pressure based laser cooling. We demonstrate that this technique provides similar atom number for lower end temperatures, provides additional cooling per scattering event and is compatible with other laser cooling methods. This method can be instrumental in bringing new exotic species and molecules to the ultracold regime.

\end{abstract}

\begin{CJK*}{UTF8}{min}

\maketitle

\end{CJK*}

The asymmetry between matter and antimatter is one of the great mysteries of modern physics. A promising avenue to better understand this asymmetry is to precisely compare spectra of hydrogen with antihydrogen, but this requires the ability to generate robust trapped samples of ultracold antihydrogen \cite{Andresen2010TrappedAntiH, Gabrielse2012TrappedAntiH, Ahmadi2017AntiH1S2S, Ahmadi2017AntiHHyperfine, Ahmadi2018AntiH1S2PLymanAlpha}. Unlike hydrogen, antihydrogen can not be cooled through evaporation \cite{Masuhara1988EvapCoolH}. Indeed, antihydrogen atoms are produced in such small numbers that collisions are rare and thermalization rates are impractical \cite{Andresen2010TrappedAntiH, Gabrielse2012TrappedAntiH}. Instead, laser cooling is needed at $\unit[121.6]{nm}$, where performances remain strongly constrained by current laser technology despite heroic efforts \cite{Gabrielse2018OL_LymanAlphaSource, Hansch2001PRL_CWLymanAlpha, Walraven_PRL1993_lasercoolingHydrogen, Donnan2013ProposalAntiHCooling}. The proposal in \citet{Wu2011AntiHydrogen} overcomes these limitations by allowing the extraction of many photon recoils of momentum per scattering event by the simple addition of an optical lattice.

This method is not limited to cooling antihydrogen \cite{Wu2011AntiHydrogen}. Related proposals have considered a similar working principle to cool and localize atoms \cite{Taieb1994SisyphusLattice, Ivanov2011SisyphusInODT} as well as to decelerate hot Yb atomic beams \cite{Ivanov2014LoadingOptLatt}. Slowing and cooling ultracold molecules \cite{DeMarco2018FDGMol, Stuhl2008MOTPolarmolecules, Isaev2016PolyatomicLaserCooling} is perhaps the most topical application for this method today. Recent striking successes in molecule cooling have relied chiefly on applying laser cooling to molecules, either by choosing species with close-to-diagonal Franck-Condon factors \cite{Barry2014MOTmolSrF, Anderegg2017CaFMOT, Collopy2018YOMOT}, or by using highly efficient laser cooling techniques \cite{Kozyryev2018BichromaticSrOH, Truppe2017CaFMOTSubDoppler}. The introduction of the present cooling method can bring better cooling efficiency and a broader class of laser cooled molecules. In turn, progress in molecule cooling is opening a plethora of ways to probe the very foundations of physics \cite{Safronova2017RevNewPhysicsAtMol, DeMille2015PhysTodayMol}. Some of the most prominent include tests for the possible variation of fundamental constants \cite{Uzan2003ReviewFundCte} and tests of the validity of fundamental symmetries \cite{Kozyryev2017TimeRevViolationPolyatomic, Tokunaga2013ChiralParity, Cahn2014ParityViolation, Hudson2011ShapeEDM, Baron2013LimitEDM, Cornell2017PRL_HfFMoleculeIon}.

A range of approaches have been devised to achieve improved performance while relaxing constraints imposed by traditional Doppler cooling techniques. For example, rapid cycling using stimulated emission can provide stronger momentum transfer without spontaneous heating or loss from non closed cycling transitions. This is demonstrated in bichromatic force cooling \cite{Soeding1997Bichromatic}, adiabatic rapid passage \cite{Metcalf2007PRA_AdiabaticRapidPassagecooling} and SWAP cooling \cite{Norcia2018TheoSWAP} but it requires intense resonant light not available at the $\unit[121.6]{nm}$ transition needed for antihydrogen. Alternatively, Sisyphus-like cooling methods \cite{Dalibard1989Sisyphus}, where kinetic energy is converted into potential energy, can function effectively even at very low pumping rates and are routinely applied to beat the Doppler temperature limit \cite{Lett1988SisyphusExp}. Examples of this approach include Zeeman-Sisyphus decelerators \cite{Fitch2016ZeemanSisyphusDecelerator} and Rydberg-Stark decelerators \cite{Hogan2008StarkRydberg3D, Hogan2009StarkRydbergMol}, where a photon excitation changes the internal state allowing a significant part of the slowing to be done by an externally applied electromagnetic field gradient.

\begin{figure}[tb]
\centering
\includegraphics[width=.98\columnwidth] {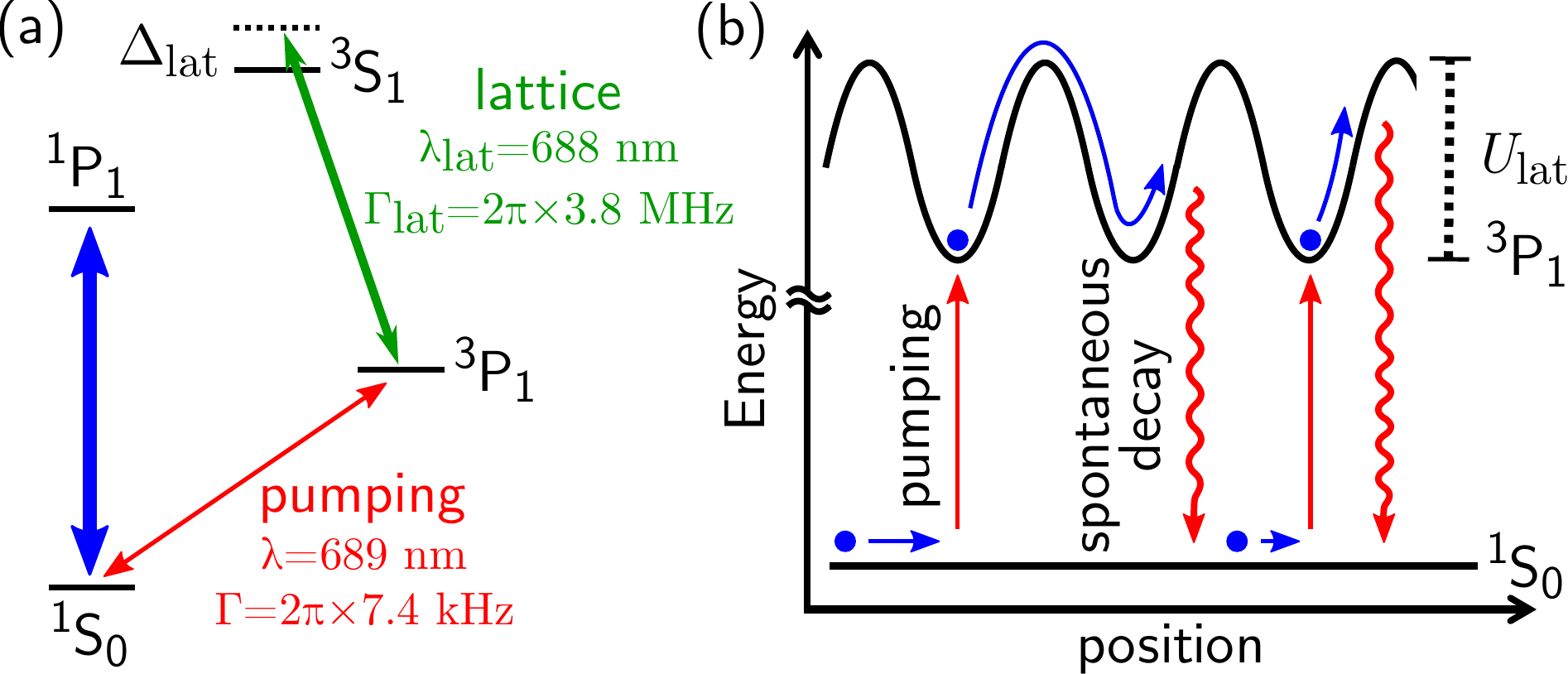}
\caption{\label{fig:Working_principle} (a) Relevant electronic levels of strontium and the two transitions used for pumping and optical lattice creation, both necessary for the SOLD. (b) Schematic of two typical cooling cycles, from pumping to spontaneous decay.}
\end{figure}

In this letter, we present a proof-of-principle demonstration of a variation on the scheme proposed by \citet{Wu2011AntiHydrogen} to laser cool antihydrogen. Without using radiation pressure, we slow a continuous stream of strontium atoms using a Sisyphus-like deceleration mechanism also described in other proposals \cite{Taieb1994SisyphusLattice, Ivanov2011SisyphusInODT, Ivanov2014LoadingOptLatt}. The method uses a 1D optical lattice acting on the excited ${^3\mathrm{P}_1}$ electronic state combined with a selective pumping mechanism that excites atoms to the lattice potential minima. We explore the performance of this technique, which we name a Sisyphus Optical Lattice Decelerator (SOLD), for various atomic beam velocities and lattice heights. Compared with the standard Zeeman slower, we show that the SOLD obtains similar fluxes but with lower temperatures. In principle, by using a deep lattice very few pumping photons can be sufficient to bring fast atoms to rest, making SOLD a good decelerator candidate for exotic species and molecules without a closed cycling transition \cite{Stuhl2008MOTPolarmolecules, Barry2014MOTmolSrF, Anderegg2017CaFMOT, Truppe2017CaFMOTSubDoppler, Collopy2018YOMOT}.

The working principle of the SOLD relies on a 3-level system coupled by two optical transitions, something ubiquitous for both atomic and molecular species. Our implementation using strontium is depicted in Fig.~\ref{fig:Working_principle}(a). An optical lattice is formed using a pair of coherent counter-propagating beams with a frequency in the vicinity of the ${^3\mathrm{P}_1} - {^3\mathrm{S}_1}$ transition. This produces a spatially modulated coupling between the ${^3\mathrm{P}_1}$ and ${^3\mathrm{S}_1}$ states and thus a spatially modulated light shift on the excited ${^3\mathrm{P}_1}$ state. The ground ${^1\mathrm{S}_0}$ state remains essentially unaffected. By applying a laser resonant with the ${^1\mathrm{S}_0} - {^3\mathrm{P}_1}$ transition, atoms can be optically pumped into the ${^3\mathrm{P}_1}$ state where they experience the force associated with the lattice potential, see Fig.~\ref{fig:Working_principle}(b). If the linewidth $\Gamma$ of the ${^1\mathrm{S}_0} - {^3\mathrm{P}_1}$ intercombination line is much smaller than the lattice height $U_{\mathrm{lat}} \gg \hbar \Gamma$, this ``pumping'' laser can be tuned to selectively address the bottom of the lattice sites. For high enough velocity $v > \lambda_{\mathrm{lat}} \Gamma$, atoms pumped into ${^3\mathrm{P}_1}$ will then climb a significant fraction of the lattice potential hills and lose kinetic energy before spontaneously decaying back to the ground state as shown in Fig.~\ref{fig:Working_principle}(b). As atoms in ${^1\mathrm{S}_0}$ continue to propagate along the lattice axis, this cooling cycle can repeat like a Sisyphus mechanism. By making the lattice very deep it is theoretically possible to remove most of the forward kinetic energy of an atom with a single cycle, as in Rydberg-Stark decelerators \cite{Hogan2008StarkRydberg3D, Hogan2009StarkRydbergMol}, and potentially within distances on the order of the lattice period. The theoretical temperature limit for this scheme is the higher of an effective Doppler temperature depending on $\Gamma$ \cite{Ivanov2011SisyphusInODT}, or the recoil temperature associated with spontaneous emission.

\begin{figure}[tb]
\centering
\includegraphics[width=.98\columnwidth]{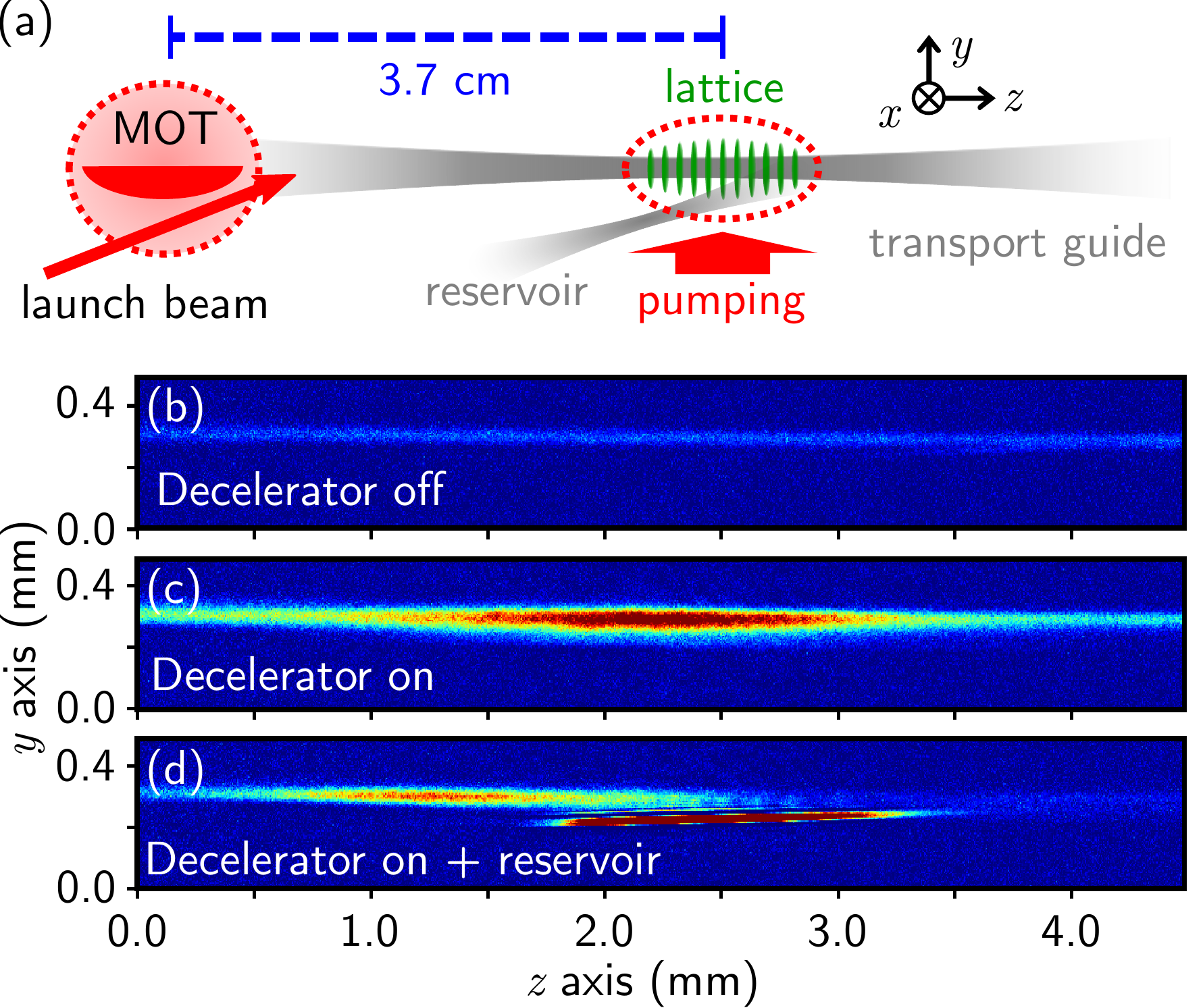}
\caption{\label{fig:Architecture_and_absorption_images} (a) Side view of the setup. (b,c,d) ${^1\mathrm{S}_0} - {^1\mathrm{P}_1}$ absorption imaging pictures of the atomic beam at the location of the decelerator, (b) without lattice, (c) with lattice, and (d) with both lattice and reservoir trap.}
\end{figure}

To demonstrate experimentally the feasibility of the SOLD, we implement the setup shown in Fig.~\ref{fig:Architecture_and_absorption_images}(a). We start with a magneto-optical trap (MOT) operating in a steady-state regime on the $\unit[7.4]{kHz}$-linewidth ${^1\mathrm{S}_0} - {^3\mathrm{P}_1}$ line, as described in our previous work (configuration ``Red MOT I'' of \cite{Bennetts2017HighPSDMOT}). We overlap this MOT with an optical dipole trap acting as a ``transport'' guide \cite{InPrep}. This 1D guide is $\sim \SI{35}{\micro\K}$ deep at the MOT location and propagates horizontally along the $z$ axis. By adding a ``launch'' beam resonant with the ${^1\mathrm{S}_0} - {^3\mathrm{P}_1}$ $\pi$ transition and pointed at the overlap between the MOT and transport guide, we outcouple MOT atoms into the guide with a well-controlled mean velocity ranging from $0.08$ to $\unit[0.25]{m.s^{-1}}$ \cite{SupplMat}. Atoms then propagate along the transport guide for $\sim \unit[3.7]{cm}$ until they reach the decelerator region.

We produce a 1D lattice potential with a pair of counter-propagating laser beams whose frequency is blue-detuned by $\Delta_{\mathrm{lat}} \approx 2 \pi \times \SI{30}{\giga\Hz}$ from the ${^3\mathrm{P}_1} - {^3\mathrm{S}_1}$ transition. The lattice beams cross the transport guide at a shallow angle of $\SI{6}{\degree}$, overlapping the atomic beam for about $\SI{3.4}{\milli\metre}$. Optical ``pumping'' from the $ {^1\mathrm{S}_0}$ to ${^3\mathrm{P}_1}$ state is provided by illuminating the atoms from the radial direction. Pumping laser beams are $\unit[15]{kHz}$ red detuned from the $\pi$ transition and their combined intensity corresponds to a saturation parameter of $\sim 1$. In addition to pumping, these beams provide an optical molasses effect, which brings the atoms' radial temperature to $\sim \SI{2}{\micro \K}$. Importantly, there is no near-resonant light capable of slowing atoms in the $z$ axis in the absence of the SOLD optical lattice.

We operate the decelerator on a guided atomic beam continuously fed by the MOT, with a homogeneous axial density across the full field of view of our imaging system, see Fig.~\ref{fig:Architecture_and_absorption_images}(b). When the lattice is switched on, the density in the overlap region between the atomic and lattice beams sharply increases, suggesting an accumulation of slowed atoms, as shown in Fig.~\ref{fig:Architecture_and_absorption_images}(c). Without either lattice beam or with a large ($\unit[160]{MHz}$) frequency difference between the two lattice beams, this feature vanishes. Fig.~\ref{fig:Architecture_and_absorption_images}(c) also shows that some atoms travel completely across the lattice region due to incomplete slowing or by diffusion. Note that our slowing mechanism is fully compatible with a steady-state apparatus, and we perform our measurements after reaching steady state.

\begin{figure}[tb]
\centering
\includegraphics[width=.98\columnwidth]{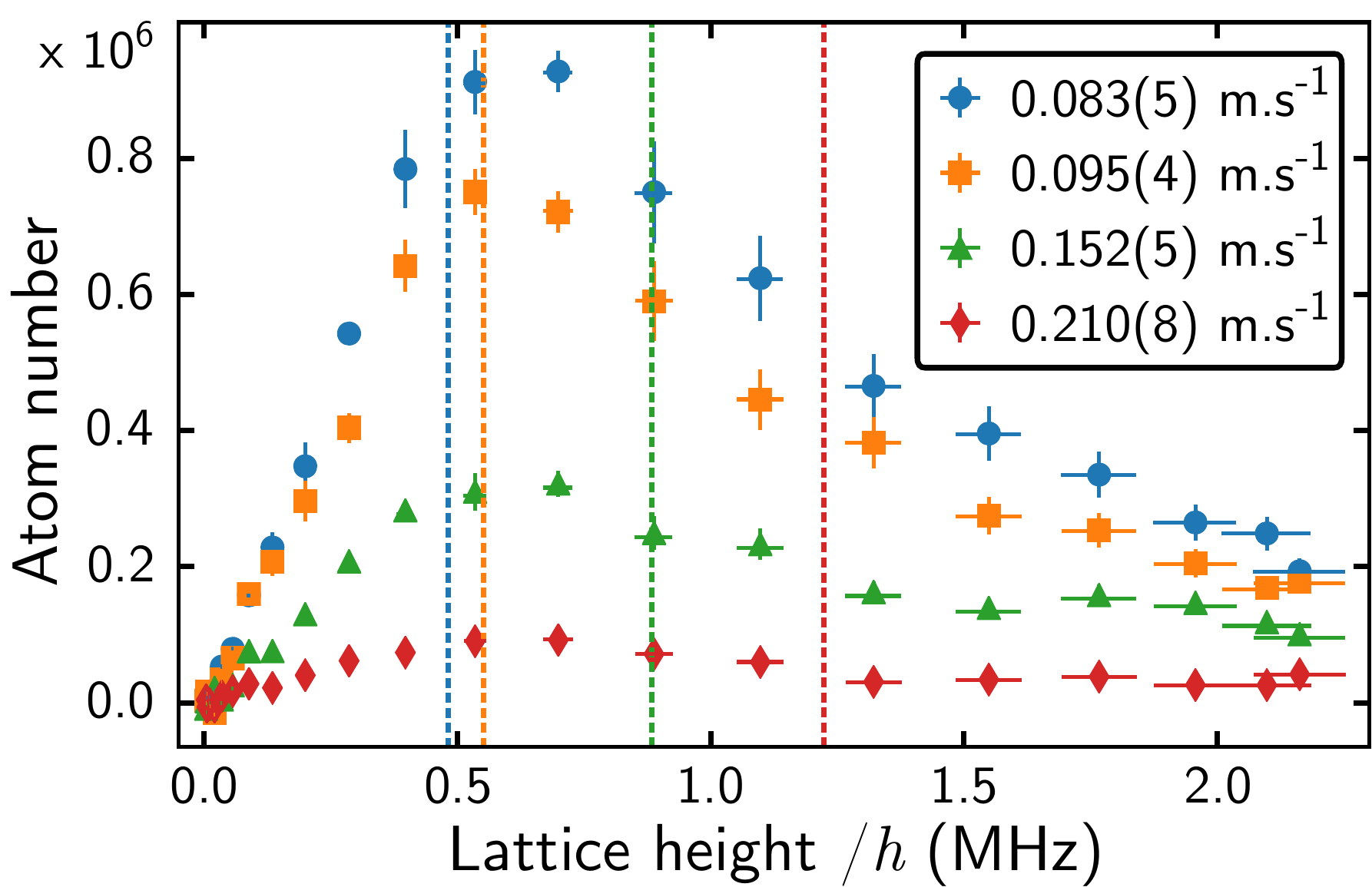}
\caption{\label{fig:Cooling_vs_velocity} Measured steady-state number of atoms slowed by the SOLD and loaded in the reservoir, for varying lattice height and four different initial velocities. The dash vertical lines give the criterion of eq.~(\ref{eq:Scat_Crit}) for $m = 1$. The vertical error bars represents standard errors from binned data points. The horizontal error bars origin is described in \cite{SupplMat}.}
\end{figure}

For better characterization of the SOLD, and since we are concerned about diffusion of slowed atoms, we add a second ``reservoir'' dipole trap beam. This beam crosses below the transport guide at the lattice location, with an offset adjusted to allow slow atoms to pass from the guide into the reservoir while not significantly disturbing the potential landscape of the guide. Thus, the reservoir collects slowed atoms $\unit[2]{mm}$ away from the crossing, which are ultimately stored there with the help of optical molasses, see Fig.~\ref{fig:Architecture_and_absorption_images}(a). We measure the mean velocity selectivity of the loading of this reservoir \cite{SupplMat}, which matches a Gaussian centered around zero velocity with a width $\sigma_v = \unit[0.0084(4)]{m.s^{-1}}$. We show one example of loading into this reservoir in Fig.~\ref{fig:Architecture_and_absorption_images}(d), which also exemplifies a means of atom extraction from our ultracold atom source. We show in Fig.~\ref{fig:Cooling_vs_velocity} the measured atom number loaded into the reservoir by the SOLD. The efficiency is poor for small lattices, as not enough kinetic energy is removed before atoms leave the lattice location. For increasing lattice height, we observe a clear loading optimum, followed by a slow decrease. These two features originate from the behavior of the pumping rate to ${^3\mathrm{P}_1}$.

\begin{figure}[tb]
\centering
\includegraphics[width=.98\columnwidth]{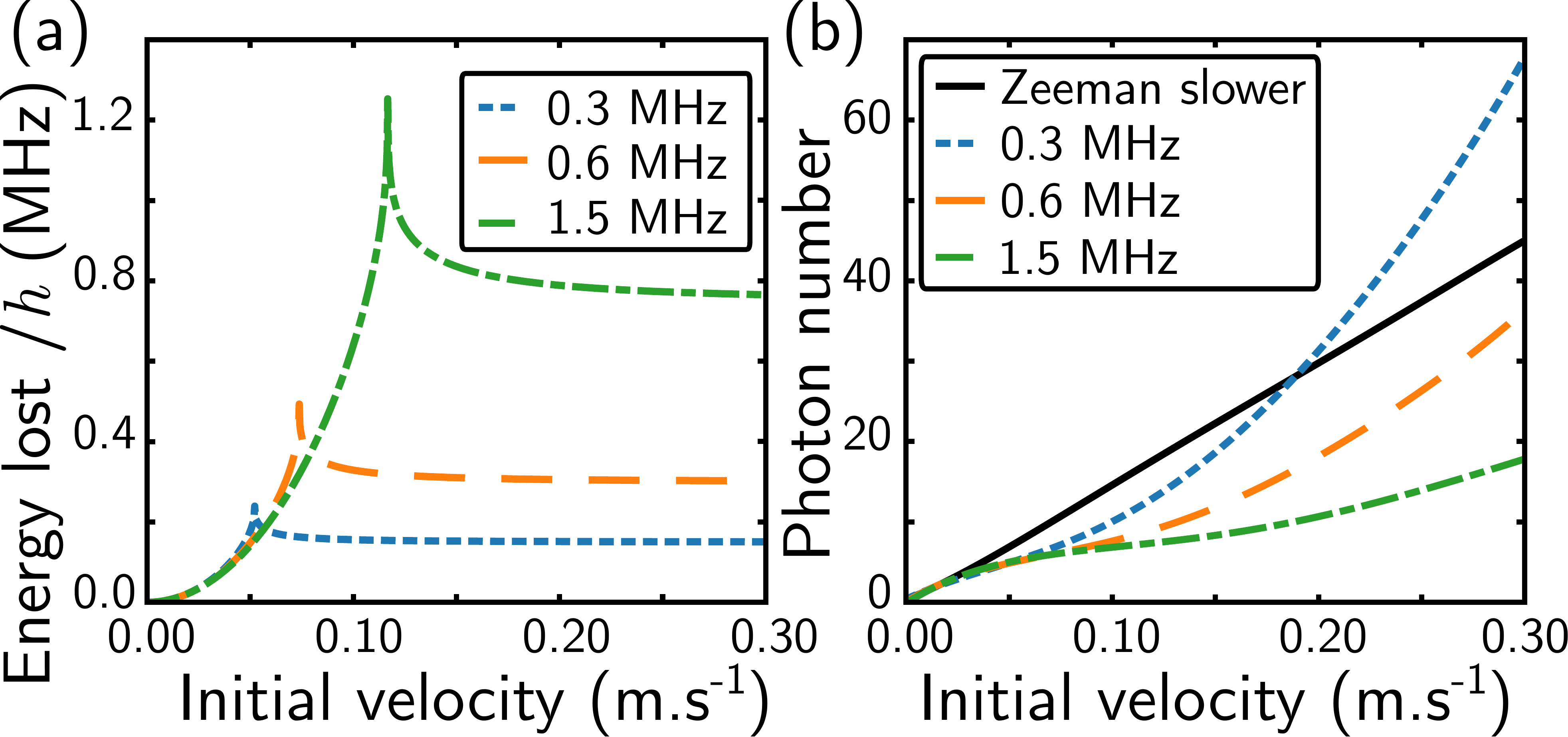}
\caption{\label{fig:Theory_results} Efficiency of our cooling scheme for varying initial velocity. (a) Average energy $E_{\mathrm{lost}}$ lost during the first cooling cycle. (b) Number of cycles/pumping photons needed for the SOLD compared with a Zeeman slower. The plain black line shows the behavior of the ZS, while the dotted, dashed and dash-dotted lines are for the SOLD with lattice heights $U_{\mathrm{lat}} = h \times \unit[0.3]{MHz}$, $\unit[0.6]{MHz}$ and $\unit[1.5]{MHz}$, respectively.}
\end{figure}

We can better understand the observed SOLD slowing efficiency by using a simple semi-classical model describing its various working regimes, which depend on the relative magnitude of the atoms kinetic energy with respect to the lattice height. Consider an atom initially pumped into the ${^3\mathrm{P}_1}$ state at the bottom of the lattice potential. In Fig.~\ref{fig:Theory_results}(a), we plot the dependence of the average energy lost per pump cycle $E_{\mathrm{lost}}$ with incoming velocity $v$ and lattice height. For high kinetic energies compared to the lattice height $\frac{1}{2} m v^2 \gg U_{\mathrm{lat}}$, atoms travel through several lattice sites and the energy lost saturates to $E_{\mathrm{lost}} \rightarrow U_{\mathrm{lat}} / 2$ provided that $v \gg \lambda_{\mathrm{lat}} \Gamma$. A striking feature of Fig.~\ref{fig:Theory_results}(a) is that $E_{\mathrm{lost}}$ exhibits an efficiency peak for $\frac{1}{2} m v^2 = U_{\mathrm{lat}}$. In this case, atoms have just enough energy to climb one lattice maximum, where they spend most of their time and are thus more likely to undergo spontaneous emission. The energy lost asymptotically reaches $E_{\mathrm{lost}} \rightarrow U_{\mathrm{lat}}$ for $v \gg \lambda_{\mathrm{lat}} \Gamma$ \cite{SupplMat}.

An important benchmark for a laser cooling technique is the average number of photons which needs to be scattered to slow atoms from some initial velocity to the final velocity allowed by this technique. In Fig.~\ref{fig:Theory_results}(b) we calculate the number of pumping photons needed to reach a kinetic energy equivalent to a temperature below $\SI{2}{\micro K}$. This temperature was arbitrarily chosen $\sim 4$ times larger than the recoil temperature associated with the ${^1\mathrm{S}_0} - {^3\mathrm{P}_1}$ transition, the relevant cooling limit in our case. For comparison with radiation pressure based laser cooling methods, we also show in Fig.~\ref{fig:Theory_results}(b) the number of photons required in the case of a Zeeman slower (ZS) \cite{Phillips1982PRL_ZeemanSlower}. The SOLD requires always less cooling photons than the ZS for a lattice height satisfying $U_{\mathrm{lat}} / h > v / \lambda_{\mathrm{lat}}$.

\begin{figure}[tb]
\centering
\includegraphics[width=.98\columnwidth]{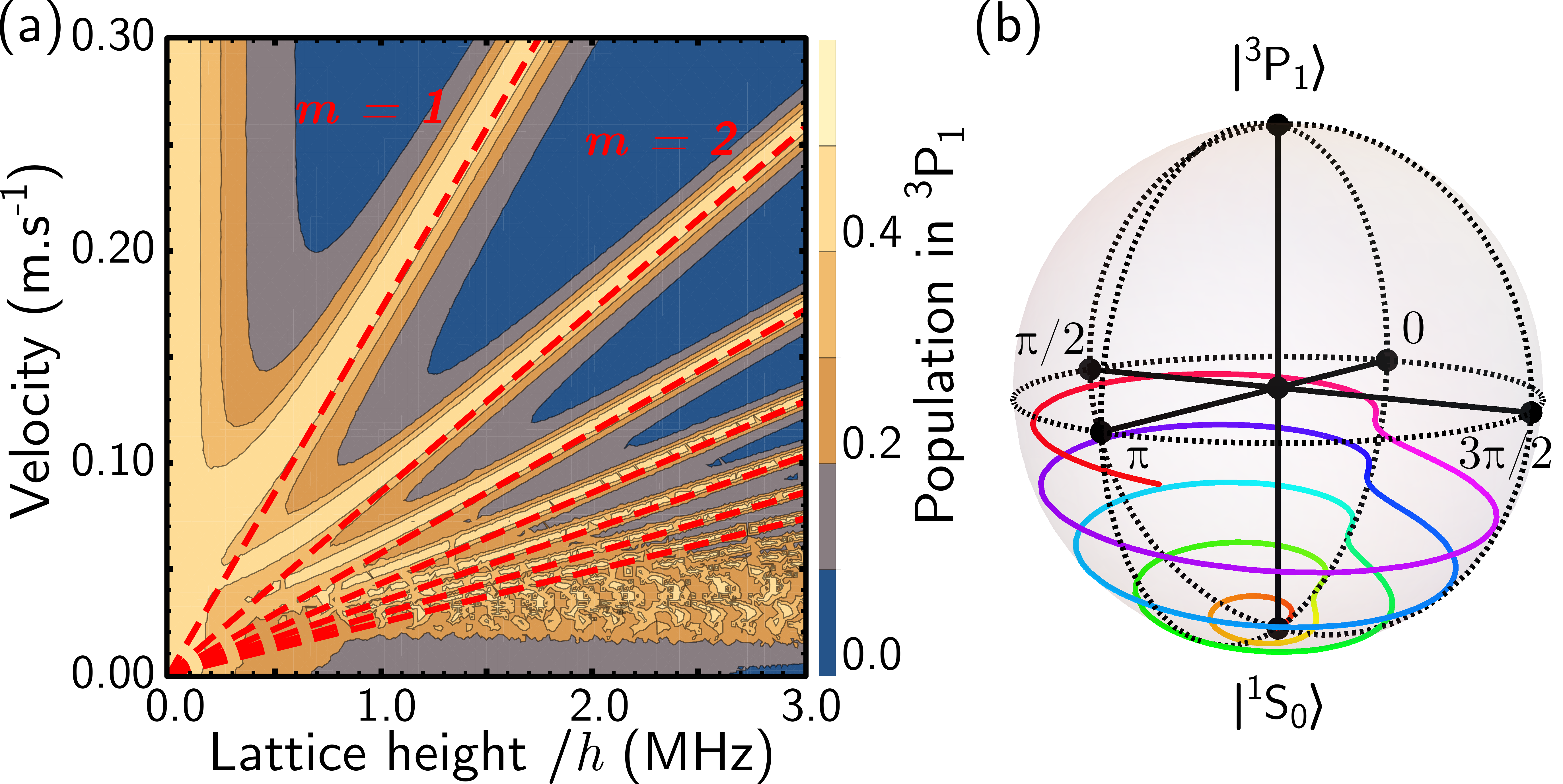}
\caption{\label{fig:Excited_population_and_Bloch_Vector} (a) Population transferred to the ${^3\mathrm{P}_1}$ state depending on the lattice height and the atom velocity. For clarity of the figure, the population is calculated for a saturation parameter of the pumping transition of $\sim 1600$ instead of the $0.1 \sim 10$ typically used. Dashed red lines show the condition of eq.~(\ref{eq:Scat_Crit}) for $m \in \{ 1...7 \}$. (b) Short time evolution of the bloch vector (for 5 ``pulses'') in the case of in-phase excitation satisfying eq.~(\ref{eq:Scat_Crit}) with $m=1$.}
\end{figure}

The SOLD ability to slow atoms with high incoming velocities is strongly dependent on the pumping rate. We model this rate by solving the optical Bloch equations for a two-level system in dependence of velocity and lattice height. Assuming a constant velocity, we numerically solve \cite{SupplMat} for the average population in ${^3\mathrm{P}_1}$, which we show in Fig.~\ref{fig:Excited_population_and_Bloch_Vector}(a). The remarkable feature in Fig.~\ref{fig:Excited_population_and_Bloch_Vector}(a) is the presence of multiple resonances where there are high pumping rates. These can be explained by in-phase multiple $\pi$-over-$N$ pulses. Indeed, only at the bottom of a lattice site is the detuning small enough to pump a significant population to ${^3\mathrm{P}_1}$. While the atoms propagate from one site to the next, the ${^1\mathrm{S}_0}$ and ${^3\mathrm{P}_1}$ states acquire different phases. Once at the next site, further population is efficiently pumped to ${^3\mathrm{P}_1}$ only if the dephasing is equal to multiples of $2 \pi$, see Fig.~\ref{fig:Excited_population_and_Bloch_Vector}(b). This criterion on the dephasing leads to the relation
\begin{equation}
\label{eq:Scat_Crit}
\frac{U_{\mathrm{lat}}}{h} = m \times \frac{4 v}{\lambda_{\mathrm{lat}}} ,
\end{equation}
with $m \in \mathbb{N}$ and $h$ the Planck constant. The optimums in loading efficiency observed in Fig.~\ref{fig:Cooling_vs_velocity} correspond mainly to the fulfillment of this criterion for the case $m = 1$. Including both the average lost energy $E_{\mathrm{lost}}$ and the pumping rate, we model the behavior of the SOLD and reproduce qualitatively the features of the experimental data \cite{SupplMat}. Moreover, we find that the criterion of eq.~(\ref{eq:Scat_Crit}) with $m = 1$ dictates the capture velocity of the SOLD, which reads $v_c = U_{\mathrm{lat}} \lambda_{\mathrm{lat}} / 4 h$. For a lattice height thus matching the atoms' velocity, the SOLD requires less photons than standard radiation pressure based laser cooling methods like the Zeeman slower.

We now compare experimentally the SOLD performance with that of a Zeeman slower. To this end, we add a laser beam counter-propagating to the transport guide, focused in the SOLD region and with a circular polarization set to address the $m_{\mathrm{J}} = -1$ Zeeman sub-state. We demonstrated in previous work that it is possible to operate a ZS on the narrow Sr intercombination line \cite{Bennetts2017HighPSDMOT}. We report in Tab.~\ref{tab:Comparison_exp_parameters} a comparison between the two slowing methods. Both give similar results for fluxes and final atom numbers, with an advantage for the ZS, which we attribute mainly to the spatial selectivity of its optical excitation. However, we observe a clear difference in the final axial temperatures $T_z$ within the reservoir, which effectively reflects the final mean velocities. For the SOLD, $T_z$ is almost as low as the radial temperature $T_{\mathrm{rad}}$ provided by the molasses cooling, whereas $T_z$ is $2.5$ times hotter for the ZS. This is because a Zeeman slower is unable to decelerate atoms to zero velocity, as they remain somewhat resonant with ZS photons and are pushed backwards. On the contrary, for the SOLD the final mean velocity is stationary in the frame of the optical lattice, whose velocity can be set at will by the frequency difference between lattice beams \cite{Salomon1996PRL_BlochOscillationinOpticalLattice, Raizen1996PRL_AcceleratingOpticalLattice, SupplMat}.

\begin{table}[tb]
\caption{Comparison of the SOLD and the Zeeman slower (ZS). The rows give steady-state atoms numbers, fluxes, $1/e$ loading times, and radial (axial) temperatures $T_{\mathrm{rad}} $ ($T_{\mathrm{z}}$). The various configurations are, in order, the SOLD in the transport guide, the SOLD plus the reservoir trap (R), the ZS plus reservoir and the combination of both techniques.}
\begin{ruledtabular}
\begin{tabular*}{\columnwidth}{@{\extracolsep{\fill}}lcccc}
	& SOLD & SOLD+R & ZS+R & SOLD+ZS+R\\
 \noalign{\smallskip} \hline \noalign{\smallskip}
 	Atom $(\times 10^6)$ & $0.78(01)$ & $0.69(01)$ & $1.87(04)$ & $2.00(10)$\\
	Flux $(\times 10^6 \, \mathrm{s^{-1}})$ &  $0.74(04)$ & $0.65(03)$ & $2.11(14)$ & $2.80(15)$ \\
	Loading $\SI{}{(ms)}$ & $705(20)$ & $625(52)$ & $434(43)$ & $507(55)$ \\
	$T_{\mathrm{rad}} \, \SI{}{(\micro\K)}$ & & $1.53(02)$ & $1.08(04)$ & $1.34(02)$\\
	$T_{\mathrm{z}} \, \SI{}{(\micro\K)}$ & & $2.30(06)$ & $5.67(94)$ & $2.59(10)$\\
\end{tabular*}
\end{ruledtabular}
\label{tab:Comparison_exp_parameters}
\end{table}

An additional difference is that, since the SOLD does not rely on radiation pressure from the pumping beam to cool, it is possible to use a much broader class of transitions than for standard laser cooling methods. It is for example possible to use the ZS beam as a pumping beam that features both spatial and velocity selectivity. The lattice, now acting on atoms in ${^3\mathrm{P}_1} \, m_{\mathrm{J}} = -1$, is the one charged with decelerating atoms to zero axial velocity. In presence of both lattice and ZS beams, we observe the best number of atoms in the reservoir, while keeping the low temperature $T_z$ due to the SOLD, see Tab.~\ref{tab:Comparison_exp_parameters}.

Let us now turn to considerations for further applications of this cooling scheme. Firstly, it is clear from Fig.~\ref{fig:Excited_population_and_Bloch_Vector}(a) that, at high velocities, pumping rates are low unless the lattice height matches the conditions of eq.~(\ref{eq:Scat_Crit}). This can be dealt with by temporal modulation of the lattice intensity, which varies the resonance locations. Secondly, for lattices much higher than the transport guide depth, we observe a clear spread of the atomic beam out of the guide. This is due both to the radial anti-confinement from the blue-detuned lattice beams and the slight angle between lattice and transport beams. A red-detuned lattice could remedy this by confining the atoms radially, but this will make correctly tuning the pumping frequency dependent on the lattice intensity. Thirdly, if the lattice detuning $\Delta_{\mathrm{lat}}$ is insufficient, atoms in the ${^3\mathrm{P}_1}$ state can be optically pumped by the lattice light to ${^3\mathrm{S}_1}$ \cite{SupplMat}, where they can decay to the metastable ${^3\mathrm{P}_0}$ and ${^3\mathrm{P}_2}$ states, introducing a need for repumping \cite{Dinneen1999Repump707679Sr}. Finally, the initial velocities decelerated in this proof of principle are low compared with several applications of interest, in part due to the small lattice deceleration region used. In the proposal of \citet{Wu2011AntiHydrogen}, the lattice is $\unit[78.5]{MHz}$ high and the capture velocity is set to $v_{c,\bar{H}} \approx \unit[25]{m.s^{-1}}$. It would take about 20 photon scattering events for the SOLD to bring antihydrogen close to the recoil limit, which, as seen in Fig.~\ref{fig:Theory_results}(b), is similar to the numbers demonstrated in this work. If we consider the same velocity and lattice height for much heavier elements such as strontium, the number of photons required by the SOLD is about 1700, much lower than the ZS case of 3700.

To summarize, we experimentally demonstrate the feasibility of a Sisyphus-type decelerator first proposed in a slightly different form to laser cool antihydrogen \cite{Wu2011AntiHydrogen, Taieb1994SisyphusLattice, Ivanov2011SisyphusInODT, Ivanov2014LoadingOptLatt}. The efficiency of this decelerator is characterized in the steady-state regime both experimentally and theoretically in dependence of incoming atom velocities and lattice heights. We determine the capture velocity of this technique, and compare the SOLD with a Zeeman slower, where we show axial temperatures a factor of $2.5$ lower. With both techniques combined, we profit from both the ZS spatial/velocity selectivity and the SOLD end temperatures. Finally, we briefly consider some improvements and applications to the case of antihydrogen. In order to efficiently slow atoms the SOLD method requires only three simple requirements; a three level system, selective pumping in a lattice with $U_{\mathrm{lat}} \gg \hbar \Gamma$, and an initial velocity satisfying $v > \lambda_{\mathrm{lat}} \Gamma$. Such simple requirements can be fulfilled by many systems where laser cooling to the ultracold regime remains a challenge, such as new exotic species and (polyatomic) molecules \cite{Stuhl2008MOTPolarmolecules, Barry2014MOTmolSrF, Anderegg2017CaFMOT, Truppe2017CaFMOTSubDoppler, Collopy2018YOMOT, Isaev2016PolyatomicLaserCooling, Kozyryev2018BichromaticSrOH}. Moreover, by careful choice of the time sequence for the lattice velocity and intensity, a pulsed version of the SOLD could bring an atom wavepacket to any desired velocity while scattering only a handful of photons.

\textit{Note added} Recently, we became aware of related Sisyphus cooling in optical tweezers \cite{Cooper2018TweezerSr}.

\begin{acknowledgments}
B.P. thanks the NWO for funding through Veni grant No. 680-47-438. We thank the NWO for funding through Vici grant No. 680-47-619 and the European Research Council (ERC) for funding under Project No. 615117 QuantStro. C.-C. C. thanks support from the MOE Technologies Incubation Scholarship from the Taiwan Ministry of Education.

C.-C. C. and S.B. contributed equally to this work.
\end{acknowledgments}

%%\bibliographystyle{apsrev4-1-etal}
%%\bibliographystyle{unsrtnat} 
%\bibliography{SisyphusLattice}


\newpage

\section{Supplemental material}

\subsection{Sisyphus Optical Lattice Accelerator}

\begin{figure}[b]
\centering
\includegraphics[width=.98\columnwidth]{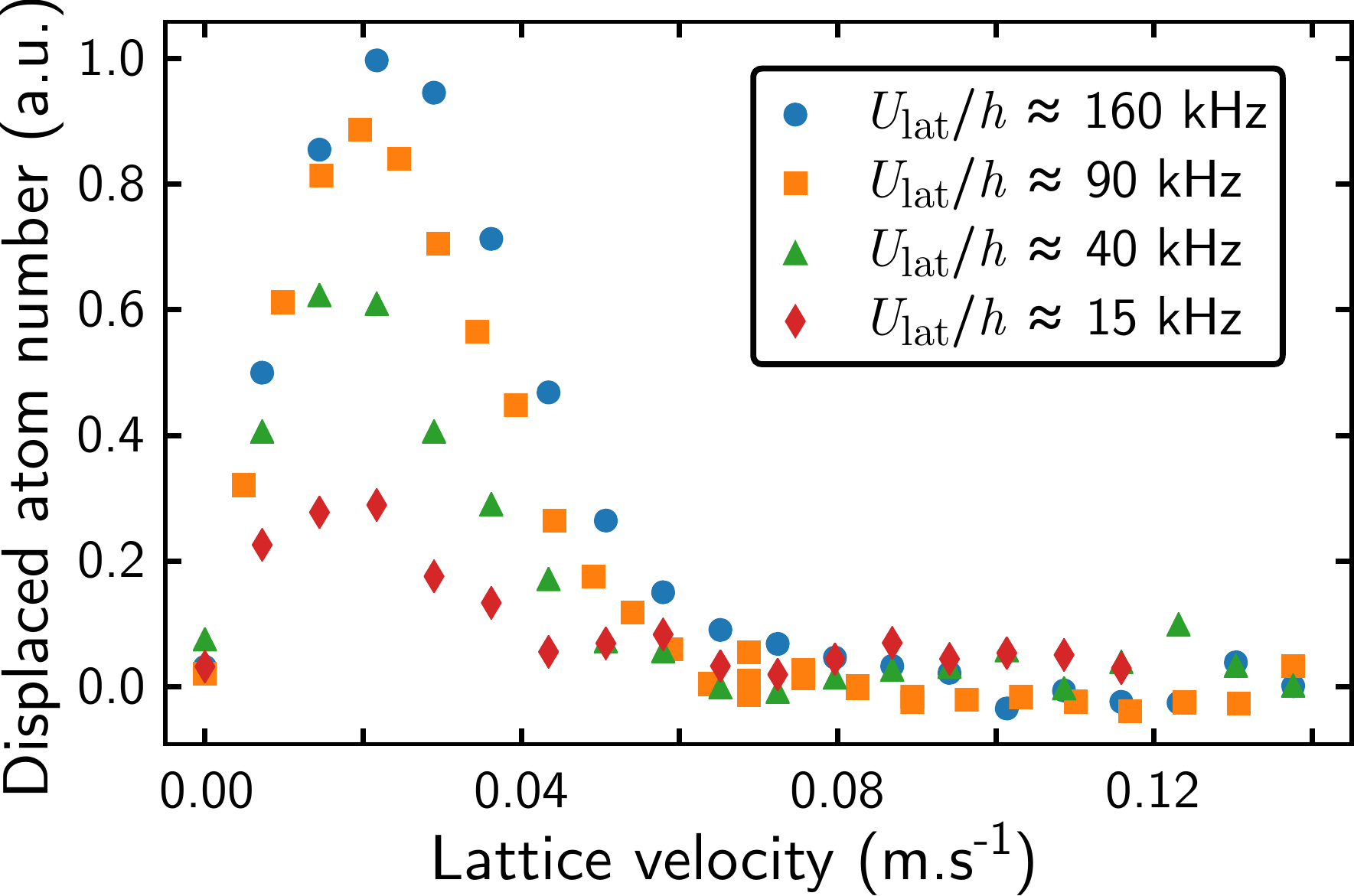}
\caption{Acceleration of a stationary strontium cloud by a moving lattice, for various lattice heights. The abscissa gives the lattice velocity and the ordinate, in arbitrary units, is proportional to the fraction of atoms in the moving frame measured after $\SI{100}{\micro s}$ of acceleration followed by $\SI{20}{\milli s}$ of evolution.}
\label{fig:SM_Displaced_atoms_vs_lattice_velocity}
\end{figure}

The SOLD deceleration scheme brings atoms ultimately to zero mean velocity in the reference frame of the lattice. By applying a small frequency difference between two lattice beams, a lattice will move at a well-controlled velocity \cite{Salomon1996PRL_BlochOscillationinOpticalLattice, Raizen1996PRL_AcceleratingOpticalLattice}. This implies that the SOLD can ideally decelerate or accelerate atoms to any desired velocity. We test this using a $\SI{1.53(2)}{\micro K}$ stationary cloud produced by loading a MOT into a dipole trap, at the location of the lattice. We shine both lattice and pumping light onto this cloud for $\SI{100}{\micro s}$, and after $\SI{20}{\milli s}$ observe the number of atoms in a displaced cloud corresponding to the moving lattice frame. The results are shown in Fig.~\ref{fig:SM_Displaced_atoms_vs_lattice_velocity}. We observe an increase in the displaced fraction with lattice height, which we attribute to the increase in energy $\sim U_{\mathrm{lat}} / 2$ given to the atoms for each scattering event. We also observe an optimal lattice velocity for a given lattice height, which roughly corresponds to our model criterion of eq.~(\ref{eq:Scat_Crit}) with $m=1$. The variation in the location of these efficiency peaks is more visible in Fig.~\ref{fig:SM_Displaced_atoms_vs_lattice_velocity} than in Fig.~\ref{fig:Cooling_vs_velocity}, because here the SOLD is pulsed for a short duration instead of operating in the steady-state regime, so the effects of each resonance corresponding to eq.~(\ref{eq:Scat_Crit}) are more pronounced. Note that due to the initial size of the cloud and its location with respect to the lattice, our estimation of the effective lattice depth is much rougher than for the data of Fig.~\ref{fig:Cooling_vs_velocity}.

\begin{figure}[b]
\centering
\includegraphics[width=.98\columnwidth]{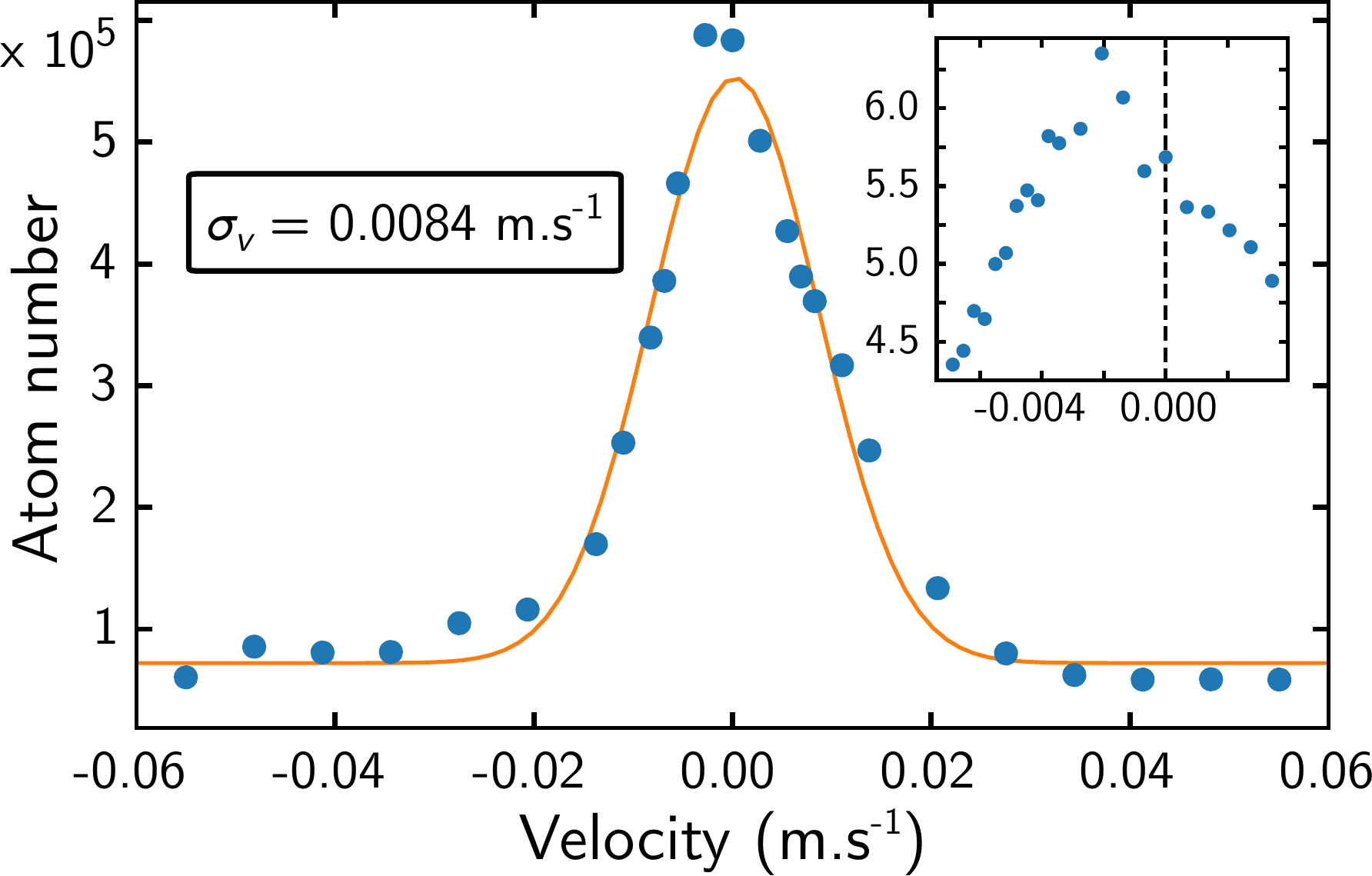}
\caption{Velocity selectivity of the reservoir loading, measured by varying the lattice velocity. The line is a Gaussian fit of the data with width $\sigma_v$. The inset shows the same type of measurement on a much narrower velocity range, highlighting the center velocity of about $v_{\mathrm{R}} \sim \unit[-0.002]{m.s^{-1}}$.}
\label{fig:SM_reservoir_loading_vs_lattice_velocity}
\end{figure}

We can also use the moving lattice to characterize our reservoir dipole trap. The loading of this reservoir is both sensitive to the mean velocity of atoms and to the location they end up when reaching zero mean velocity. We characterize the velocity acceptance of the reservoir by varying the frequency difference between the two lattice beams. The loading efficiency of the reservoir depending on the lattice velocity is shown in Fig.~\ref{fig:SM_reservoir_loading_vs_lattice_velocity}. It can be fitted by a Gaussian whose width is $\sigma_v = \unit[0.0084(4)]{m.s^{-1}}$, centered at $v_{\mathrm{R}} \sim \unit[-0.002]{m.s^{-1}}$. This slight departure from zero velocity can be explained by the orientation of the reservoir relative to the guide, which favors the loading of atoms that move backward. We include this measured velocity selectivity of the reservoir in our model of the SOLD.

\subsection{SOLD Model}

Here we give a description of our model of the SOLD that is an extended version of the description given in the main text.

In order to model our cooling scheme in an insightful way, we split the problem into two parts: the average energy lost per cooling cycle and the pumping rate. We then use both results to simulate the time evolution of the atoms' velocity.

\subsubsection{Energy lost}

We begin with a study of the energy lost due to the presence of the lattice. We assume that the atoms are optically pumped into the ${^3\mathrm{P}_1}$ state at the bottom of the lattice and we solve the differential equation for the motion $z(t)$ along the lattice propagation axis:
\begin{equation}
\label{eq:Propagation}
\frac{1}{2} m v_0^2 = U_{\mathrm{lat}} \sin^2 k_{\mathrm{lat}} z + \frac{1}{2} m \left(\frac{d z}{d t}\right)^2,
\, z(t=0) = 0, 
\end{equation}
with $m$ and $v_0$ being respectively the mass and the initial velocity of the atom. $U_{\mathrm{lat}}$ is the lattice depth and $k_{\mathrm{lat}} = \frac{2 \pi}{\lambda_{\mathrm{lat}}}$ is the wave vector of the lattice light with wavelength $\lambda_{\mathrm{lat}}$. The solution of this equation can be written in terms of the Jacobi amplitude $J_A$:
\begin{equation}
\label{eq:propagation_solution}
z(t) = \frac{1}{k_{\mathrm{lat}}} J_A \left( k_{\mathrm{lat}} v_0 \, t, \frac{2 U_{\mathrm{lat}}}{m v_0^2}\right) .
\end{equation}

Since the process relies on spontaneous emission towards ${^1\mathrm{S}_0}$, we determine the average energy lost $E_{\mathrm{lost}}(U_{\mathrm{lat}} ,\, v_0)$ by integrating the lattice height explored for a duration set by the natural linewidth $\Gamma$ of the ${^1\mathrm{S}_0} - {^3\mathrm{P}_1}$ transition,
\begin{equation}
\label{eq:energy_lost}
E_{\mathrm{lost}} = \Gamma \int_{0}^{\infty} e^{-\Gamma t} \, U_{\mathrm{lat}}  \sin^2 (k_{\mathrm{lat}} z(t)) \, dt.
\end{equation}

In Fig.~\ref{fig:Theory_results}(a), we show the evolution of $E_{\mathrm{lost}}$ for several lattice heights and depending on the incoming velocity. We observe that for high incoming kinetic energies compared to the lattice height $\frac{1}{2} m v_0^2 \gg U_{\mathrm{lat}}$, the energy lost $E_{\mathrm{lost}}$ saturates. In this case, atoms travel through several lattice sites, and their propagation tends to $z(t) \rightarrow \frac{1}{k_{\mathrm{lat}}} J_A \left( k_{\mathrm{lat}} v_0 \, t, 0\right) = v_0 \, t $. Equation (\ref{eq:energy_lost}) gives the relation $E_{\mathrm{lost}} \rightarrow \frac{U_{\mathrm{lat}}}{2} / \left(1+ \left( \frac{\Gamma}{ 2 k_{\mathrm{lat}} v_0}\right)^2 \right) $. In our experiment $v_0 \gg \lambda_{\mathrm{lat}} \Gamma$, so the average energy lost saturates to $U_{\mathrm{lat}} / 2$. One striking feature of Fig.~\ref{fig:Theory_results}(a) is that the energy lost exhibits a sharp resonance for $\frac{1}{2} m v_0^2 = U_{\mathrm{lat}}$, where cooling is the most efficient. In this case, atoms have just enough energy to climb on top of the first lattice hill, so they spend most of their time at this location, which makes them more likely to spontaneously emit there and therefore to lose most of their kinetic energy. Indeed, the explored lattice height becomes $U(t) \rightarrow U_{\mathrm{lat}} \tanh^2  ( k_{\mathrm{lat}} v_0 \, t )$, which for $v_0 \gg \lambda_{\mathrm{lat}} \Gamma$ gives an average energy lost reaching asymptotically $E_{\mathrm{lost}} \rightarrow U_{\mathrm{lat}}$. Let us note that, in contrast to Ref.~\cite{Wu2011AntiHydrogen}, which relies also on a spatial modulation of $\Gamma$, the effective rate of spontaneous emission in our case is higher on lattice hills only because of the increased time atoms spend there.

\subsubsection{Pumping rate}

We now examine the pumping rate in dependence of the incoming velocity and lattice height. We solve the optical Bloch equation for a two-level system corresponding to the ${^1\mathrm{S}_0}$ and ${^3\mathrm{P}_1}$ states, coupled by the pumping laser with Rabi frequency $\Omega$. The time-dependent Schr\"odinger - von Neumann equation for the density operator $\rho$ is
\begin{equation}
\label{eq:OBE_density_operator}
\frac{d \rho}{d t} = -\frac{i }{\hbar}  [H , \, \rho]  + L ,
\end{equation}
with $\hbar$ the reduced Planck constant, $L$ the usual term to account for the spontaneous emission due to $\Gamma$, and with the Hamiltonian $H$ written as:
\begin{equation}
\label{eq:OBE_hamiltonian}
H = 
\begin{pmatrix}
0 & \Omega / 2 \\
\Omega / 2 & U_{\mathrm{lat}} \sin^2 (k_{\mathrm{lat}} v_0 \,t ) 
\end{pmatrix} .
\end{equation}

We numerically solve eq.~(\ref{eq:OBE_density_operator}) with time, starting with all the population in ${^1\mathrm{S}_0}$ at $t = 0$. For this calculation, we assume a constant velocity $v_0$, which is valid for $\frac{1}{2} m v_0^2 \gg U_{\mathrm{lat}}$. After a variable time, the $(\Omega,\, U_{\mathrm{lat}},\, v_0)$-dependent solution for the excited population reaches a steady-state only slightly perturbed by the time-dependent detuning produced by travelling within the lattice. Averaging over this small perturbation, we get the population in the ${^3\mathrm{P}_1}$ state shown in Fig.~\ref{fig:Excited_population_and_Bloch_Vector}.

\begin{figure}[tb]
\centering
\includegraphics[width=.98\columnwidth]{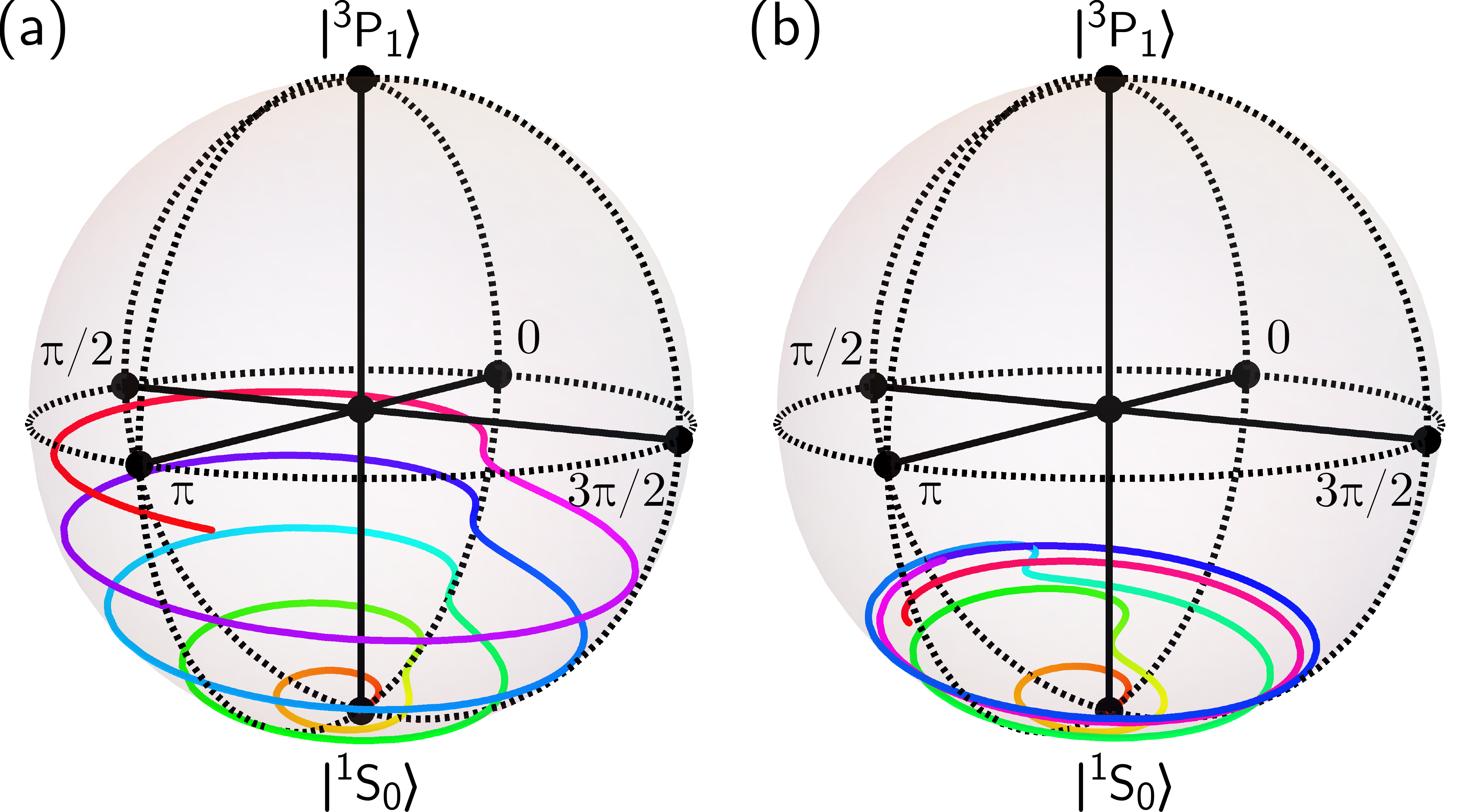}
\caption{\label{fig:SM_Bloch_vector} Evolution of the Bloch vector on the Poincar\'e sphere, shortly after the application of the SOLD. Atoms begin in the $^1\mathrm{S}_0$ state at the location of a lattice site. The saturation parameter of the pumping to $^3\mathrm{P}_1$ is set for clarity to about 60, and the lattice height is $h \times \unit[600]{kHz}$. The velocity in (a) $v_0 = \unit[0.1035]{m.s^{-1}}$ is such that the accumulated phase during the travel time between two sites is close to $\Phi = 2 \pi$, while in (b) where $v_0 = \unit[0.09]{m.s^{-1}}$ this condition is not met.}
\end{figure}

The remarkable feature in this figure is the presence of sharp lines where the excited population is the highest. These can be simply explained by looking at the evolution of the Bloch vector associated with $\rho$, displayed in Fig.~\ref{fig:SM_Bloch_vector}. When the atoms are not located at the bottom of a lattice site, the detuning is so strong that effectively the scattering rate vanishes, so the population distribution remains roughly constant for timescales short relative to $\Gamma$. Due to the dephasing between both states, the Bloch vector then evolves mainly horizontally, until the atoms reach the location of the bottom of the next lattice site. If at that moment the dephasing amounts to a multiple of $2 \pi$, then the effective pumping pulses add constructively, and the steady-state excited population is high. This leads to the resonance lines in Fig.~\ref{fig:Excited_population_and_Bloch_Vector}.

We can give a simple quantitative criterion for the positions of these lines. The phase accumulated during the propagation through one lattice period is $\Phi = \Delta \, T$, with $T = \frac{\lambda_{\mathrm{lat}}}{2 v_0}$ the propagation time and $\Delta$ the dephasing, taken as the average detuning due to the lattice, giving $\Delta = 2\pi \, \frac{1}{h} \, \frac{U_{\mathrm{lat}}}{2}$, with $h$ the Planck constant. The condition $\Phi = m \times 2 \pi$ (with $m \in \mathbb{N}$) gives the relation:
\begin{equation}
\label{eq:SM_Scat_Crit}
\frac{U_{\mathrm{lat}}}{h} = m \times \frac{4 v_0}{\lambda_{\mathrm{lat}}} .
\end{equation}
This criterion is shown as dashed red lines for $m \in \{ 1, \, ..., \, 7 \}$ in Fig.~\ref{fig:Excited_population_and_Bloch_Vector}.

\subsubsection{Overall evolution}

\begin{figure}[tb]
\centering
\includegraphics[width=.98\columnwidth]{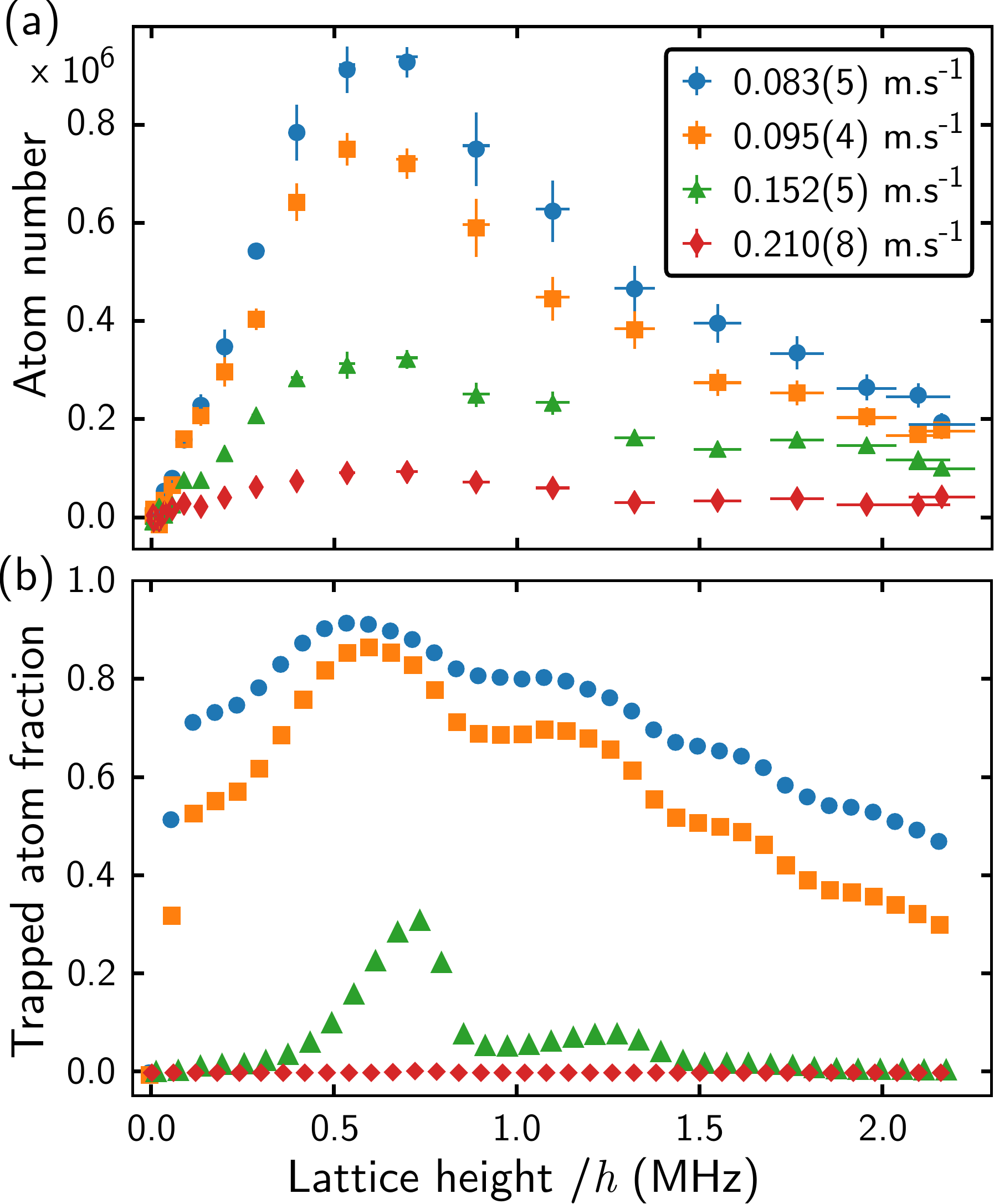}
\caption{\label{fig:SM_Comparison_theory_data} Comparison between (a) the experimental data already shown in Fig.~\ref{fig:Cooling_vs_velocity} and (b) the results of the theoretical model (see text) for the same initial mean velocities. }
\end{figure}

In order to model the complete behavior of the SOLD, we solve classically the evolution of the atoms' velocity $v$ with time, under an effective force $F(U_{\mathrm{lat}}, \, v) = - \Gamma \times \rho_{^3\mathrm{P}_1}(U_{\mathrm{lat}}, \, v) \, E_{\mathrm{lost}}(U_{\mathrm{lat}}, \, v)$. We carry out this calculation for a packet of atoms whose velocity distribution follows a (1D) Boltzmann distribution corresponding to the temperature of our MOT of $\SI{6}{\micro K}$ summed with an offset corresponding to the measured mean velocity given by the launch beam. The capture probability in our reservoir is determined by the velocity-dependent efficiency extracted from the measurement shown in Fig.~\ref{fig:SM_reservoir_loading_vs_lattice_velocity}, corresponding to a Gaussian function with a width $\sigma_v = \unit[0.0084]{m.s^{-1}}$. We thus simulate the time evolution of the loaded population in the reservoir depending on the lattice height, for the four mean starting velocities shown in Fig.~\ref{fig:Cooling_vs_velocity}. In Fig.~\ref{fig:SM_Comparison_theory_data} we compare the results from this model with our experimental data.

We see a good qualitative agreement concerning the overall behavior with both lattice height and starting mean velocity. In particular, the locations of the optimums of loading efficiency are well reproduced by our model. These correspond to the case when the starting mean velocity $v_0$ verifies the criterion of eq.~(\ref{eq:Scat_Crit}) (with $m = 1$). Indeed, in that case atoms are efficiently pumped to the ${^3\mathrm{P}_1}$ state, and lose typically a significant amount of energy $U_{\mathrm{lat}} / 2$. After spontaneous emission, their velocity is much lower and atoms are in the $(U_{\mathrm{lat}}, \, v)$ region where the density of lines for $m \geq 2$ is high. They are therefore very likely to keep decelerating efficiently. On the contrary, for high velocity $v_0$, in the region $0 \ll \frac{U_{\mathrm{lat}}}{h} \ll \frac{4 v_0}{\lambda_{\mathrm{lat}}}$, atoms will not get pumped to ${^3\mathrm{P}_1}$. Our model is thus able to estimate the capture velocity $v_c$ of the SOLD, which is given by 
\begin{equation}
\label{eq:Capture_velocity}
v_c = \frac{U_{\mathrm{lat}} \lambda_{\mathrm{lat}}}{4 h} .
\end{equation}

Let us note that our model makes several approximations. Indeed the results of the calculations shown in Fig.~\ref{fig:SM_Comparison_theory_data}(b) are given for one particular evolution time $t= \unit[1.4]{ms}$ that has been chosen for best match with the steady-state experimental data. Since no decay mechanism has been added in the model, the final loading would be with unity efficiency. This chosen deceleration time is rather short, because in this case the saturation parameter of the ${^1\mathrm{S}_0} - {^3\mathrm{P}_1}$ transition is set to $\sim 320$, for which the calculations suffer less numerical errors compared to more realistic, lower saturation parameters. Nonetheless, the theory always exhibits the same overall behavior no matter the value of the saturation parameter. Another limitation of our model is that no selection criteria have been chosen for the position of atoms, whereas they must be in the vicinity of the crossing between the transport guide and reservoir to be loaded. Similarly, atoms expelled from the guide by the barrier formed by the blue detuned lattice and the effects of the lattice's slight angle with the guide are not taken into account. Finally, the constant velocity approximation made when solving the optical Bloch equations is not valid for $\frac{1}{2} m v_0^2 \leq U_{\mathrm{lat}}$. To obtain a better quantitative agreement, Monte-Carlo simulations could be a straightforward option for further studies.

\subsection{Atomic beam velocities}

In order to characterize the dependence of the SOLD efficiency with incoming atom velocity, we need a measurement of the mean atom velocity within the transport guide before entering the lattice region. In the absence of the SOLD, we will assume the mean velocity to be constant throughout the lattice region, as the potential provided by the transport guide in this location is engineered to be flat in the axial direction.

\begin{figure}[tb]
\includegraphics[width=.98\columnwidth]{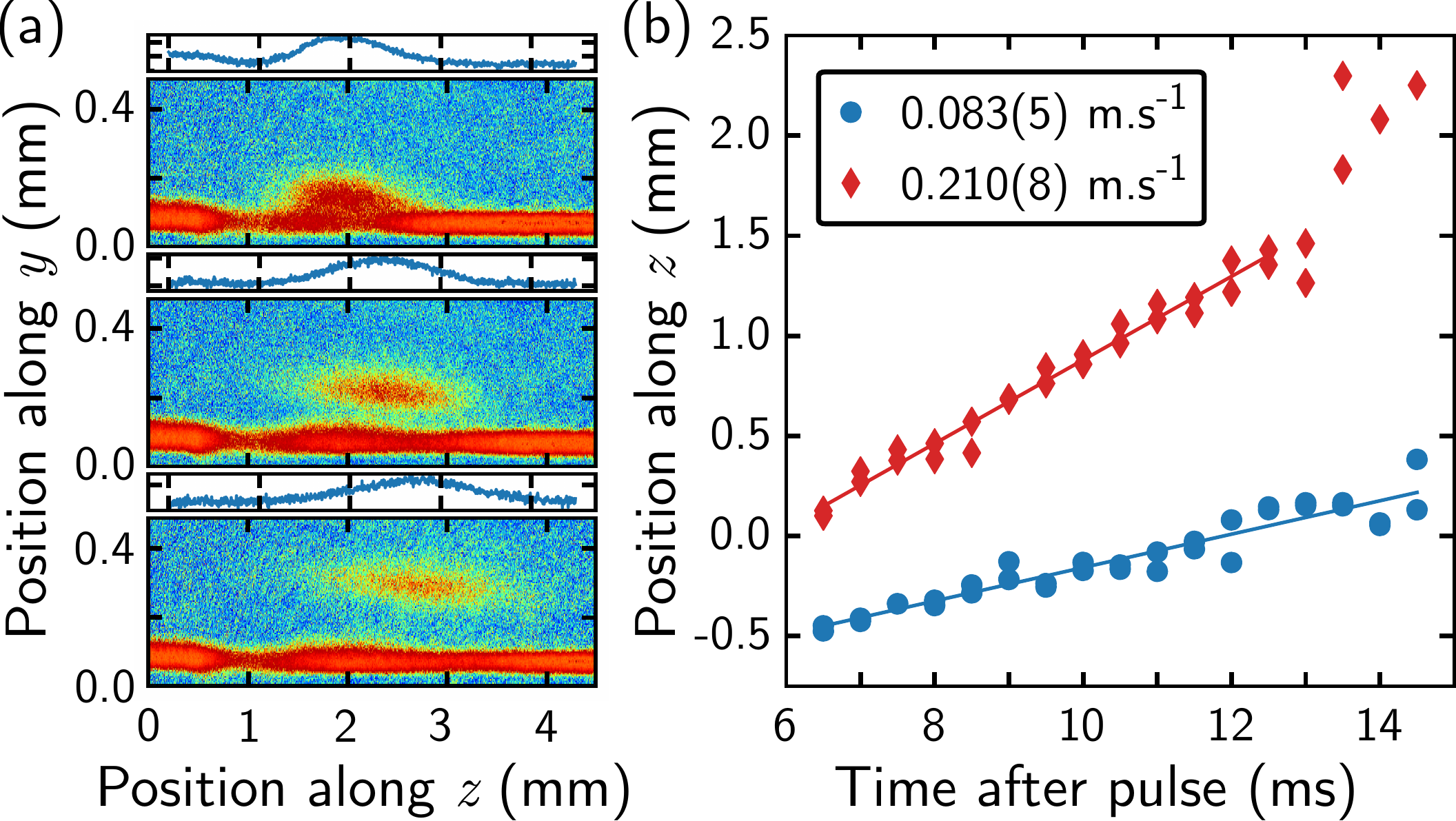}
\caption{\label{fig:SM_Atomic_beam_velocity} Measure of the atom beam velocity. (a) Absorption images and respective density profiles of clouds of atoms ejected from the transport guide, taken $\unit[5]{ms}$ (top), $\unit[8]{ms}$ (middle) and $\unit[10]{ms}$ (bottom) after the ejection pulse. The pulse is applied at $z \approx \unit[1]{mm}$. (b) Evolution of the position along the $z$ axis of two ejected clouds with different mean velocities. The lines are fits from which we extract the velocities.}
\end{figure}

We measure the atom velocity arriving at the location of the lattice by two methods. The first is to eject a burst of atoms out of the transport guide using a pulse of light resonant with the ${^1\mathrm{S}_0} - {^1\mathrm{P}_1}$ transition. The pulse lasts $\unit[1]{ms}$ and the laser beam propagates horizontally perpendicular to the atomic beam. We assume that spontaneously emitted photons are equally distributed in all directions during the ejection process, so that the on-axis mean velocity is not affected by the light absorption. By examining the propagation of the packet of ejected atoms, we can infer the mean velocity, see Fig.~\ref{fig:SM_Atomic_beam_velocity}. We were only able to measure the mean velocity with high accuracy, as measurements of the axial distribution or temperatures of these ejected clouds were limited by low signal to noise. Rough estimates were however consistent with the measured MOT temperature of $\SI{6}{\micro K}$.

We use a second method as a cross-check for our determination of the atom mean velocity. It relies on the measurement of the loading in the transport guide region. Without the SOLD applied, we measure the steady-state atom number in the transport guide in the region corresponding to the extent of the lattice. We can extract the linear density $\rho_{\mathrm{lin}}$ in this region, if assumed to be homogeneous. By also assuming the mean velocity $v_0$ to be constant, we have the following relation for the steady-state incoming and outgoing flux:
\begin{equation}
\label{eq:Flux_and_density}
F_{\mathrm{lat,out}} = F_{\mathrm{lat,in}} =\rho_{\mathrm{lin}} \times v_0 .
\end{equation}

We thus need to measure $F_{\mathrm{lat,out}} = F_{\mathrm{lat,in}}$, which we get from the loading curves of the MOT and the transport guide. The rate equations for both MOT and transport guide atom numbers, $N_{\mathrm{MOT}}$ and $N_{\mathrm{TG}}$, are 
\begin{equation}
\label{eq:Rate_equation}
\begin{cases}
\frac{d N_{\mathrm{MOT}}}{d t} = F_{\mathrm{in}} - \beta_{\mathrm{TG,in}} \, N_{\mathrm{MOT}}  - \beta_{\mathrm{MOT,loss}} \, N_{\mathrm{MOT}} \\
\\
\frac{d N_{\mathrm{TG}}}{d t} =  \eta_{\mathrm{TG}} \,  \beta_{\mathrm{TG,in}} \, N_{\mathrm{MOT}} - \beta_{\mathrm{TG,out}} \, N_{\mathrm{TG}}
\end{cases} ,
\end{equation}
where $F_{\mathrm{in}}$ is the flux coming into the MOT, see Ref.~\cite{Bennetts2017HighPSDMOT}. The loss rate $\beta_{\mathrm{MOT,loss}}$ describes atoms lost from the MOT, while $\beta_{\mathrm{TG,in}}$ describes the rate of atoms coupled into the beginning of the transport guide. The efficiency $\eta_{\mathrm{TG}}$ accounts for the losses between the start and the end of the guide. Last, the rate $\beta_{\mathrm{TG,out}}$ describes atoms leaving the lattice region. We thus have $\beta_{\mathrm{TG,out}} \, N_{\mathrm{TG}}(t=\infty) = F_{\mathrm{lat,out}} = F_{\mathrm{lat,in}}$. From fitting the loading curves of both MOT and transport guide with the solutions of eq.~(\ref{eq:Rate_equation}), we extract the values of each $\beta$, $F$ and $\eta$ parameters, and ultimately get the value of $v_0$. We found a good agreement between the velocities derived from both methods, and we use the first one to determine the values provided in the main text.

\subsection{Losses toward {$^3\mathrm{S}_1$}}

\begin{figure}[tb]
\centering
\includegraphics[width=.98\columnwidth]{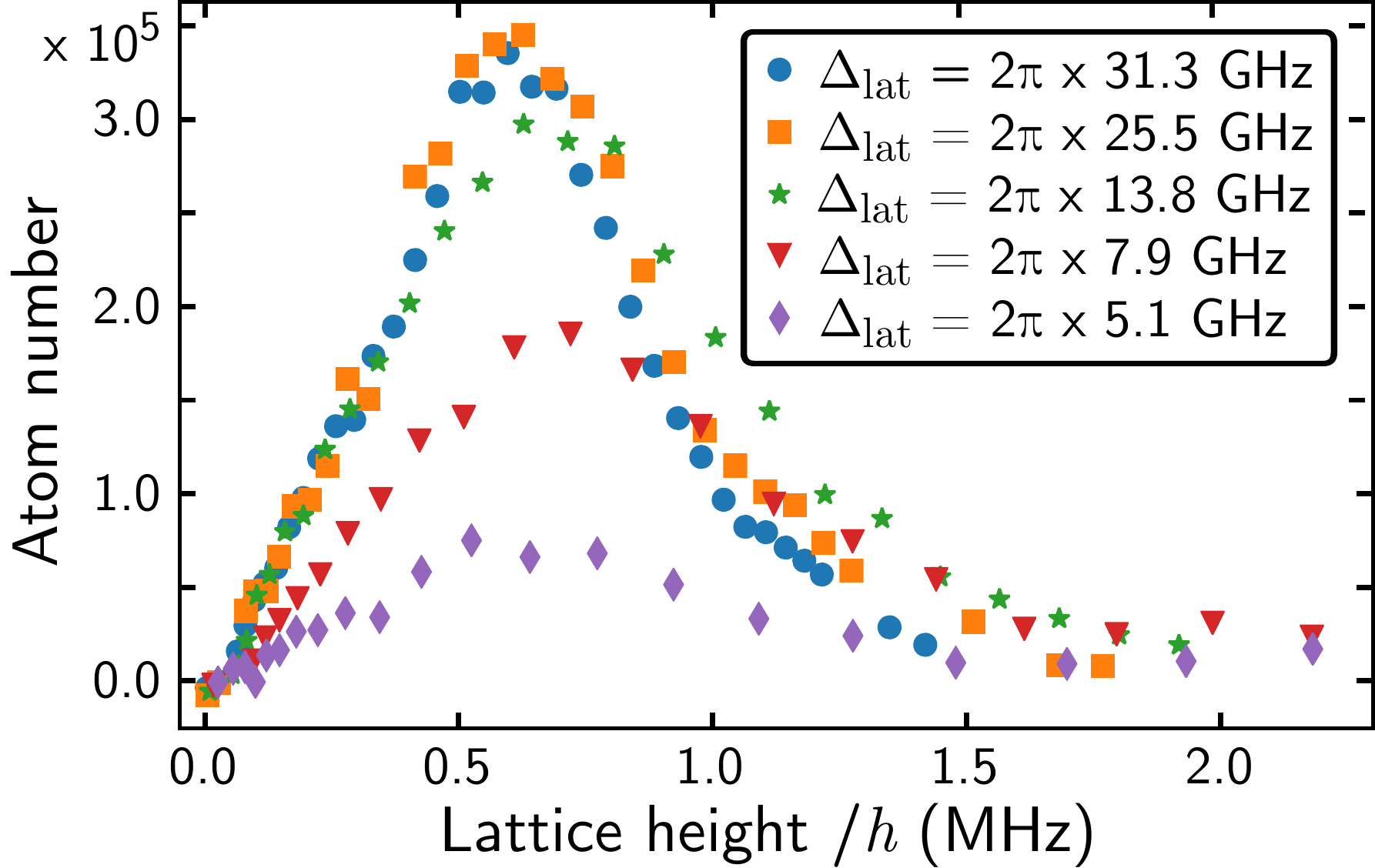}
\caption{\label{fig:Efficiency_with_optical_pumping} Effect of optical pumping by the lattice light to the ${^3\mathrm{S}_1}$ state. The data show the number of atoms loaded into the reservoir as a function of the lattice height for various detunings of the lattice laser from the ${^3\mathrm{P}_1} - {^3\mathrm{S}_1}$ transition. }
\end{figure}

One limitation of our decelerator is the optical pumping by the lattice light of atoms in the ${^3\mathrm{P}_1}$ state to the ${^3\mathrm{S}_1}$ state. If such optical pumping occurs, atoms can decay from ${^3\mathrm{S}_1}$ to the metastable ${^3\mathrm{P}_0}$ and ${^3\mathrm{P}_2}$ states and exit the cooling cycle. Fig.~\ref{fig:Efficiency_with_optical_pumping} shows, for several lattice laser detunings $\Delta_{\mathrm{lat}}$, the effect of optical pumping to ${^3\mathrm{S}_1}$ depending on the lattice height. For detunings a few $\unit{GHz}$ away from the ${^3\mathrm{P}_1} - {^3\mathrm{S}_1}$ transition we see a clear reduction of the atom number slowed and captured in the reservoir. For detunings above $\unit[20]{GHz}$, the efficiency seems to converge toward a unique curve, indicating no significant optical pumping.

A repumping scheme such as the one used in Ref.~\cite{Dinneen1999Repump707679Sr} can optically pump atoms back to ${^3\mathrm{P}_1}$. Apart from a few additional photon recoils, this method should not significantly affect the slowing process, providing the repumping time is short compared to the propagation of the atoms along the lattice. Another simple method is to detune the lattice laser frequency further away from the ${^3\mathrm{P}_1} - {^3\mathrm{S}_1}$ transition, while adapting its power to keep the same lattice height. Aside from the data of Fig.~\ref{fig:Efficiency_with_optical_pumping}, we operate at a lattice detuning of $\Delta_{\mathrm{lat}} \approx 2 \pi \times \unit[30]{GHz}$, for which optical pumping is negligible.

\subsection{Lattice height determination}

An accurate determination of the lattice height is essential for the characterization of the SOLD. The potential of a 1D lattice acting on the ${^3\mathrm{P}_1}$ state is given by
\begin{equation}
\label{eq:Lattice_polarizability}
U(z) = - \frac{1}{2\epsilon_{0} c}\,\alpha_{i} \, I(z) = - \frac{4P}{\pi\epsilon_{0} c\ w^{2}_{0}} \,\alpha_{i}\, \sin^2 k_{\mathrm{lat}} \, z,
\end{equation}
where $\alpha_{i}$ is the dynamic dipole polarizability of the $^3\mathrm{P}_1$ state, $P$ is the power of each lattice beam, and $w_0$ is their waist. In the two-level approximation,
\begin{equation}
\label{eq:Two_levels_polarizability}
\alpha_{i} \approx \frac{3 \epsilon_{0} \lambda_{\mathrm{lat}}^3 }{8 \pi^2}  \frac{\Gamma_{\mathrm{eff}}}{\Delta_{\mathrm{lat}}} ,
\end{equation}
where $\epsilon_0$ is the vacuum permittivity. The approximation is valid because the lattice laser detuning $\Delta_{\mathrm{lat}}$ is only a few tens of GHz. The effective rate $\Gamma_{\mathrm{eff}} = \eta \, A_{{^{3}\mathrm{P}_{1}} - {^{3}\mathrm{S}_{1}}}$ is the effective transition rate for the $5s5p\, {^{3}\mathrm{P}_{1}} - 5s6s\, {^{3}\mathrm{S}_{1}}$ transition, with $\eta = 1/2$ due to the lattice laser polarization. The relative uncertainties of the parameters contributing to the determination of the lattice height are listed in Tab.~\ref{tab:Uncertainty}. All parameters and their uncertainties are determined experimentally, except for the transition rate, which we derive from literature in the following manner.

\begin{table}[tb]
\caption{Relative uncertainties on the relevant parameters used to calculate the lattice height for the $^3\mathrm{P}_{1}$ state.}
\label{tab:Uncertainty}
\begin{ruledtabular}
\begin{tabular*}{\columnwidth}{@{\extracolsep{\fill}}lc}
	& uncertainty \\
 \noalign{\smallskip} \hline \noalign{\smallskip}
 	Lattice beam power $P$ &  $\pm\unit[3.0]{\%}$ \\
   	Lattice beam waist $w_0$  & $\pm\unit[1.4]{\%}$ \\
   	Lattice frequency detuning $\Delta_{\mathrm{lat}}$  & $\pm\unit[0.1]{\%}$ \\
    Total transition rate $A_{{^3\mathrm{P}_{\mathrm{J}}} - {^3\mathrm{S}_1}}$   & $\pm\unit[1.0]{\%}$ \\
    \noalign{\smallskip}
\hline\noalign{\smallskip}
Total uncertainty& $\pm\unit[4.2]{\%}$
\end{tabular*}
\end{ruledtabular}
\end{table}

The ${^3\mathrm{P}_{\mathrm{J}}} - ^3\mathrm{S}_{1}$ manifold transition rates are known accurately to the percent level. In Ref.~\cite{Safronova2013BBRClockSr}, \textit{ab-initio} calculated matrix elements for the relevant contributing transitions, together with experimental transition energies, are used to evaluate the $^{3}\mathrm{P}_{0}$ polarizability. Constraints from both the measurement of the magic wavelength at $\unit[813]{nm}$ and the dc polarizability are then imposed to fine-tune the dominant matrix element terms, in order to agree with the experimental values. The fine-tuning of these matrix elements does not exceed $\unit[1.1]{\%}$, and the theoretical transition rate for ${^{3}\mathrm{P}_{1}} - {^{3}\mathrm{D}_{1}}$ calculated from these matrix elements agrees with the measured value \cite{Nicholson2015Clock10PowMinus18} to within $\unit[0.2]{\%}$. 

From the dipole matrix elements of $5s5p\, {^{3}\mathrm{P}_{0}} - 5s6s\, ^{3}\mathrm{S}_{1}$ calculated in Ref.~\cite{Safronova2013BBRClockSr}, we determine the value of $A_{{^{3}\mathrm{P}_{1}} - {^{3}\mathrm{S}_{1}}}$. To this end, we calculate the branching ratios of the $^{3}\mathrm{S}_{1}$ state to the three $5s5p\, ^3\mathrm{P}_\mathrm{J}$ fine structure states using Wigner 6-j symbols. We take into account the fine structure splitting of the states, which is on the order of a few $\unit[100]{cm^{-1}}$, and apply the frequency dependent correction factors to the branching ratios. The resulting branching ratios are from $^3\mathrm{S}_1$ to $(^3\mathrm{P}_0, \, ^3\mathrm{P}_1, \, ^3\mathrm{P}_2) = (\unit[12.02]{\%}, \, \unit[34.71]{\%}, \, \unit[53.27]{\%})$. Using these ratios, we derive the transition rate $A_{{^{3}\mathrm{P}_{1}} - {^{3}\mathrm{S}_{1}}} = \unit[2.394(0.024) \times 10^7]{s^{-1}}$.  

\end{document}